\documentclass[journal]{IEEEtran}
	\usepackage{shellesc}
	\usepackage[utf8]{inputenc}
	\usepackage[T1]{fontenc}
	\usepackage{tikz}
	\usepackage{pgfplots}
	\pgfplotsset{compat=newest}
	\usetikzlibrary{external}
	\usepackage{pgfplotstable}
	\usepgfplotslibrary{groupplots}	
    \usepackage{cite}
    \usepackage{amsmath,amssymb,amsfonts}
    \usepackage{mathtools}
    \usepackage{algorithmic}
    \usepackage{booktabs}
    \usepackage{multirow}
    \usepackage{tabularx}
    \usepackage{graphicx}
    \usepackage{textcomp}
    \usepackage{color}
    \usepackage{glossaries}
    \usepackage{enumitem}
    \usepackage{xcolor}
	\usepackage{siunitx}
	\usepackage{float}
	\usepackage{placeins} 
	\usepackage{mathtools, stackengine}
	\usepackage{soul} 
	
	\usepackage[flushleft]{threeparttable}
	
    \DeclareSIUnit{\belmillii}{Bi}
    \DeclareSIUnit{\belmilliwatt}{Bm}
    \DeclareSIUnit{\dBm}{\deci\belmilliwatt}
    \DeclareSIUnit{\dBi}{\deci\belmillii}
    \usepackage{url}
	\usepackage[numbers]{natbib}
	\usepackage{caption}
	\usepackage{subcaption}
	\usepackage{comment}
	\usepackage{hyperref} 

	\usepackage{import}
	\graphicspath{{./figures/}}

	\usepackage{nicefrac}

\makeglossaries%
\newacronym{iot}{IoT}{Internet of Things}
\newacronym{rss}{RSS}{received signal strength}
\newacronym{snr}{SNR}{signal-to-noise ratio}
\newacronym{pl}{PL}{path loss}
\newacronym{los}{LoS}{line-of-sight}
\newacronym{per}{PER}{packet error ratio}
\newacronym{rf}{RF}{radio frequency}

\def\BibTeX{{\rm B\kern-.05em{\sc i\kern-.025em b}\kern-.08em
    T\kern-.1667em\lower.7ex\hbox{E}\kern-.125emX}}

\newcolumntype{C}{>{\centering\arraybackslash}X}
\newcolumntype{R}{>{\raggedleft\arraybackslash}X}
\newcolumntype{L}{>{\raggedright\arraybackslash}X}

\begin{document}

\title{%
RadioWeaves for efficient connectivity: analysis and impact of constraints in actual deployments%
}
\author{Liesbet Van der Perre, Erik G. Larsson, Fredrik Tufvesson, Lieven De Strycker, Emil Björnson, Ove Edfors%
	    \thanks{L. Van der Perre and L. De Strycker are with the Department of Electrical Engineering, KU Leuven, Belgium; L. Van der Perre, F. Tufvesson, and O. Edfors are with the Department of Electrical and Information Theory, Lund University, Sweden. E. G. Larsson and E. Björnson are with the Department of Electrical Engineering, Linköping University, Sweden  (e-mail: \{Liesbet.Vanderperre, Lieven.DeStrycker\}@kuleuven.be, \{Erik.G.Larsson, Emil.Bjornson\}@liu.se \{Fredrik.Tufvesson, Ove.Edfors\}@eit.lth.se) 
} 
}

\maketitle
\begin{abstract}
We present a new type of wireless access infrastructure consisting of a fabric of dispersed electronic circuits and antennas that collectively function as a massive, distributed antenna array. We have chosen to name this new wireless infrastructure ``RadioWeaves'' and anticipate they can be integrated into indoor and outdoor walls, furniture, and other objects, rendering them a natural part of the environment.  Technologically, RadioWeaves will deploy distributed arrays to create both favorable propagation and antenna array interaction. The technology leverages on the ideas of large-scale intelligent surfaces and cell-free wireless access. Offering close to the service connectivity and computing, new grades in energy efficiency, reliability, and low latency can be reached. The new concept moreover can be scaled up easily to offer a very high capacity in specific areas demanding so. 
In this paper we anticipate how two different demanding use cases can be served well by a dedicated RadioWeaves deployment: a crowd scenario and a highly reflective factory environment. A practical approach towards a RadioWeaves prototype, integrating dispersed electronics invisibly in a room environment, is introduced. We outline the many and diverse R\&D challenges that need to be addressed to realize the great potential of the RadioWeaves technology.
\end{abstract}


\section{Introduction}%
\label{sec:introduction}

\begin{figure}[ht]
    \centering
    \includegraphics[width=1\linewidth]{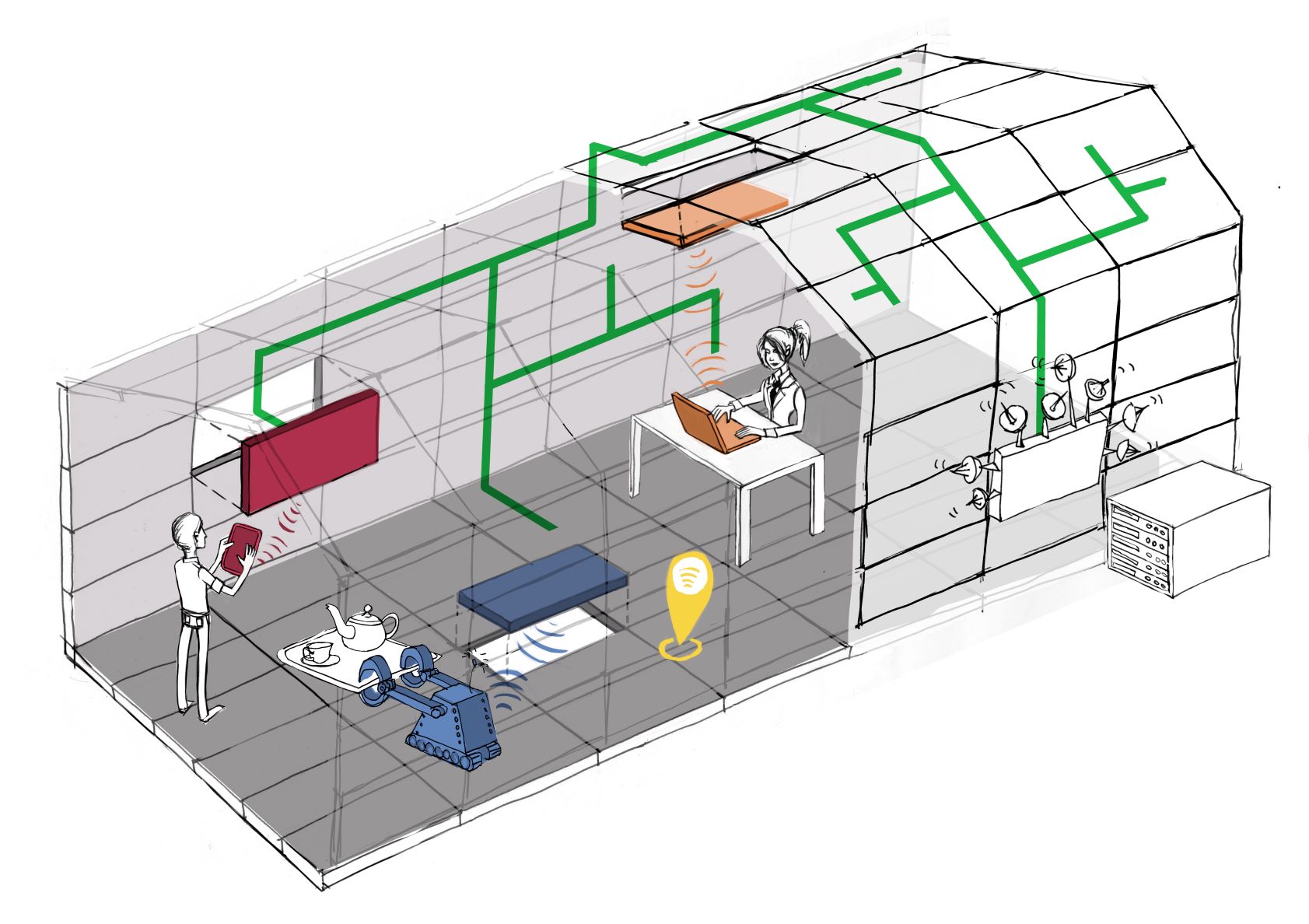}
    \caption{An artistic impression of a potential RadioWeaves implementation and operation.}%
    \label{figure:sketchRadioWeave}
\end{figure}

RadioWeaves technology presents a new wireless networking concept based on a distributed radio and computing infrastructure that will get weaved into conventional buildings and objects. In RadioWeaves, a fabric of dispersed electronic circuits and antennas collectively function as a massive, distributed antenna array. It brings both the access and the intelligence close to the devices and the applications and can hence meet requirements of services that can impossibly be supported in current networks. A RadioWeaves topology can be designed to create both favorable propagation conditions and translit and receive in the physically beneficial and efficient field of view the large antenna arrays for all devices in its coverage zone. Consequently, signals can be conditioned to operate transceivers in mild conditions. Fig.~\ref{figure:sketchRadioWeave} provides an illustration of a potential RadioWeaves implementation and operation.

The combined communication, positioning, sensing, and computation infrastructure is fit to embed more autonomy in connected systems and devices, and host intelligence close to applications. The new concept endorses locally-confined and diversity-rich communications. This  enables fast and robust exchange of information, paving the way for ultra time-critical and reliable applications. In addition, it supports the trend towards decentralisation of networking for private services.

This paper first introduces the overall RadioWeaves concept, background technologies and potential of the technology in Section~\ref{sec:concept}. We anticipate how the high capacity and ultra reliability potential of the technology could be beneficially realized in the demanding use cases of crowd scenarios and a factory environments respectively in Section~\ref{sec:analysis}. In Section~\ref{sec:impact}, we anticipate the impact of actual deployments of RadioWeaves: good, bad, and handsome. A variety R\&D challenges to be addressed to progress the novel concept to superior technology for hyper-connected environments is elaborated on in Section~\ref{sec:challenges}. Finally, we conclude on this first analysis of the RadioWeaves technology.

\section{RadioWeaves Concept}%
\label{sec:concept}
 
RadioWeaves technology will deploy a distributed topology of panels of electromagnetic elements and operate them with a cell-free wireless networking approach. It will bring the capabilities of multi-antenna systems to a next level, building upon the theoretical foundations of massive MIMO systems. A vast amount of R\&D has progressed massive MIMO technology and the first generation of implementations has demonstrated the superior capacity. However, the full potential of systems with a very large number of antenna elements in terms of energy efficiency, sensing, coverage and reliability improvements has not been demonstrated. 

RadioWeaves will blend connectivity and computational resources close to the actual application. They are deployed taking into account and actually joining forces with physical phenomena that have a major impact in real deployments:
\begin{itemize}
    \item \emph{Radio propagation.} Practical multi-path characteristics are typically far from the often assumed i.i.d. Rayleigh fading in signal processing solutions. Experimental characterisation of massive MIMO channels shows that the energy predominantly comes from one or a few specific directions, even in rather reflective environments \cite{Gunnarsson2018}. In a RadioWeaves deployment one can allocate resources to create a good link budget based on the highest energy directions. 
    \item \emph{Physical antenna arrays.} These can not be designed to transmit and receive with an omnidirectional pattern, nor do they behave as a transparent media. Real antenna arrays exhibit a non-negligible directional gain and a scan-dependent impedance~\cite{Mailloux2005}. Electromagnetic effects such as coupling between antennas and surface waves impact the effectiveness of combinations of signals sent to the antennas. Significant return losses reduce efficiency, may harm electronics, and worst case scanning blindness is encountered \cite{Pozar1984}. In RadioWeaves technology the orientation and the positioning of the surfaces and the antenna elements can be designed to avoid these problems and use the physical gain of the antenna arrays to promote desired directions.
    \item \emph{Real transceivers.} These induce distortion terms especially when "pushed" to compensate for bad transmission situations. The distortion terms will in general not be uncorrelated, and they may have undesired effects such as generating out of band radiation~\cite{larsson2018out}. A distributed array infrastructure creates a more uniform power distribution, which can reduce actual dynamic range requirements and avoid to drive PAs into nonlinear operation.
\end{itemize}

RadioWeaves technology builds upon a distributed antenna architecture implemented with intelligent surfaces integrated in the environment \cite{HU2018data} and operating in a cell-free network manner. The conceptual innovation of RadioWeaves technology thus lies in:
\begin{enumerate}
    \item The blending of radios and computing resources dispersed in the environment to offer connectivity and intelligence in the proximity of devices.
    \item The deployment of a distributed architecture of interconnected surfaces integrating radiating elements and electronics, operated with a cell-free networking approach able of dynamically achieving the best performance with the available resources. 
    \item The creation of favorable propagation conditions in the full coverage zone through an appropriate placement and orientation of antenna arrays and elements. The topology of the weave will be designed to maximize the probability of being close to one or more arrays and in a non-shadowed and favorite direction with respect to the array(s).
    \item The design of joint signal processing and scheduling solutions to offer the desired service levels at greatly improved energy efficiency, cooperating in symbiosis with the physical antenna arrays, and operating the transceivers in a gentle dynamic range.
\end{enumerate}
In the next subsections we clarify the value of distributed architectures, intelligent surfaces and cell-free networking for RadioWeaves. We further explain the opportunities to improve service levels, both for diverse connectivity needs and positioning purposes, and increase energy efficiency.

\subsection{Distributed antenna elements and arrays: concept and value for RadioWeaves} 
\begin{figure}[h]
    \centering
    \includegraphics[width=1\linewidth]{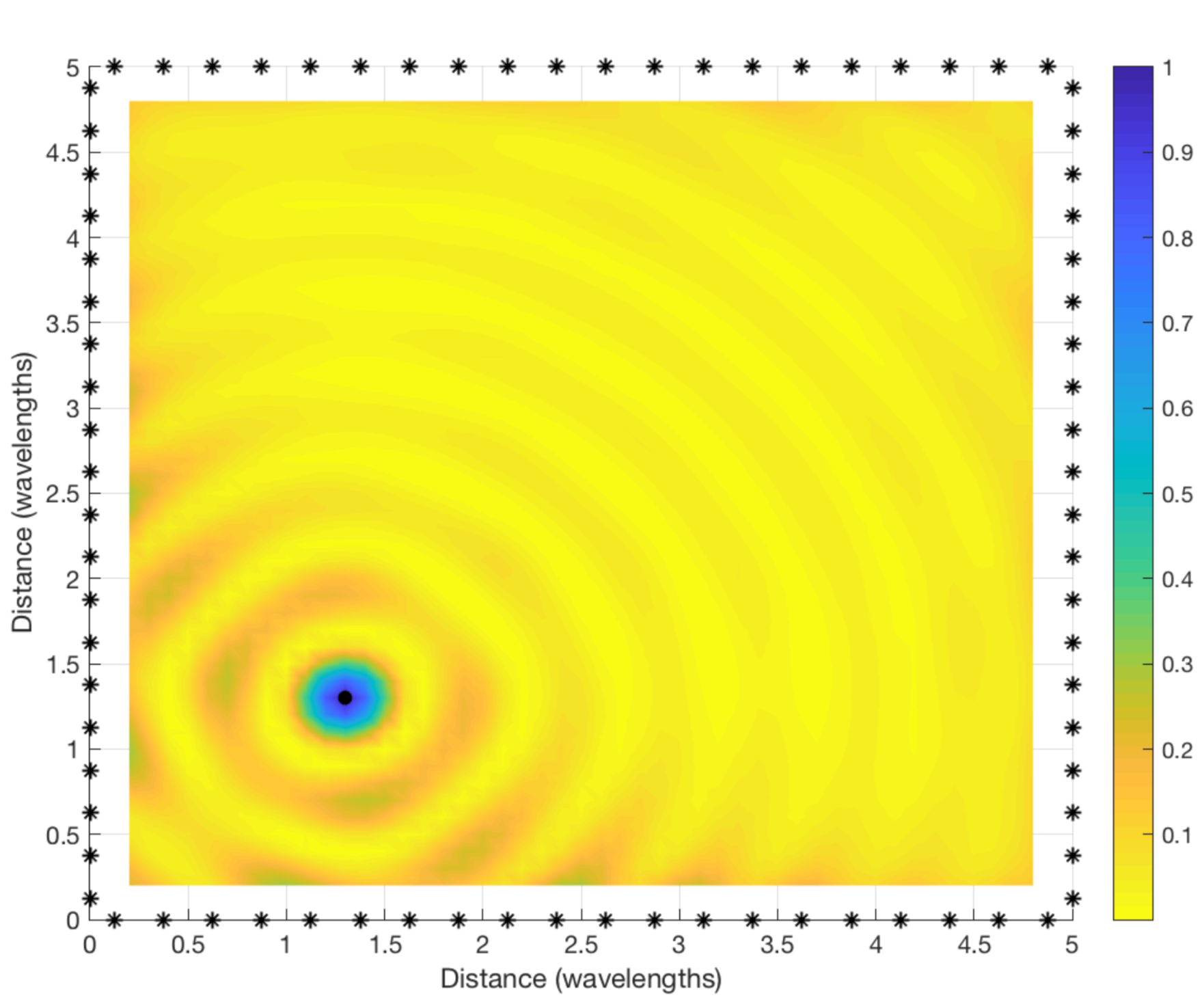}
    \caption{When transmitting from 80 antennas along the walls in a small room, coherent transmission leads to a strong signal within a ball with a diameter equal to half the wavelength.}%
    \label{figure:signal_focusing}
\end{figure}

Distributed array architectures offer several benefits compared to systems deploying one central array. They can improve coverage, reduce power consumption, and increase system capacity \cite{Clark2001, Zhou2003a}, both when considering single an multi-cell environments \cite{Choi2007}.The fundamental concept of distributed antenna systems with coherent processing is not new \cite{Shamai2001a,Zhou2003a}, and has been investigated experimentally in, e.g., \cite{Lau2012, Flordelis2013}. 
The transmission from an antenna array to a user is conventionally illustrated as a beam, but this behavior only appears for a user being in the far-field of a co-located array in a sparse propagation environment. Suppose the user is instead surrounded by antennas and in the near-field of such distributed array. Phase-aligned transmission towards the user will then result in the signal power being strong only in a small region around the receiver. This is illustrated in Fig.~\ref{figure:signal_focusing} where 80 transmit antennas are distributed along the walls of a small room. 

We notice that optimized transmission from distributed antenna elements leads to a strong signal within a region with a diameter being roughly half a wavelength\footnote{The same phenomenon can appear in co-located Massive MIMO in a rich scattering environment without dominating scatterers.}.

The benefits of a distributed array deployment in terms of spectral efficiency have also been confirmed experimentally \cite{Guevara2018}. The recently proposed ``radio stripes'' present a distributed array architecture to address capacity and coverage challenges, leveraging on a one-dimensional sequential architecture~\cite{RadioStripes}. In a RadioWeave deployment the distributed elements will be clustered in intelligent surfaces with local processing. A fabric with multiple surfaces can create a 3D illumination towards the devices. 

\subsection{Large Intelligent Surfaces: concept and value for RadioWeaves} 
The idea of large-scale intelligent surfaces is that instead of equipping a base station with a conventional antenna array, entire surfaces of natural objects or even walls in buildings can be designed to host radiating elements. A useful way of thinking of large-scale intelligent surfaces is as a continuum of antennas that extends over a large physical area, and that can transmit, receive, sense and interact with terminals and nodes in the environment. Conceptual studies have shown that this concept bears great promise both in terms of offering wireless access, and in terms of RF-based positioning \cite{HU2018data, Hu2018pos}. In RadioWeaves technology
we envision there will be interconnected surfaces, each being realized as a panel. 
The very large number of antenna elements raises considerable challenges in processing complexity and interconnect. Hence, some degree of decentralized processing within one large surface \cite{Rodriguez2019} and in coordination of different panels, will be essential. In RadioWeaves we open new degrees of freedom in the system design by establishing and dynamically operating multiple interconnected panels with different orientations and positions. This diversity on the macro-scale allows to benefit from specific propagation conditions and beneficial directions of antenna arrays. The density of array elements on a surface is a design choice where there may be good reasons for diverting from the traditional $\lambda/2$ spacing. Decreasing the density increases the surface area for a given number of array elements and may contribute to better coverage and protection against shadowing effects, while a higher density may be better from a sensing and array beamwidth performance. How to select the array element density and its effect on large intelligent surface performance in real operation is still unknown. 

\subsection{Cell-Free Networks: concept and value for RadioWeaves}


A conventional cellular network
consists of a sparse deployment of elevated base stations that each serves the users that surround them using 
high transmit power. The cell-free paradigm turns the situation around \cite{Interdonato2018a}: By deploying a large number of distributed antennas, each user will instead be surrounded by antennas that each use relatively low power. A device is simultaneously served by multiple access points surrounding this device, and this set of access points can be different for each device. There are hence no cells or cell boundaries~\cite{Ngo2017b} in contrast to a conventional small-cell scheme, whereby
each user is served by a dedicated AP. A dense spreading of access point antennas will result in a significant smaller average distance between a user and the closest antennas. The performance variations within the coverage area can consequently be greatly reduced as compared to cellular networks.
The same effect cannot be achieved by densifying a cellular network \cite{Ngo2017b,Nayebi2017a}, following the fact that the transmitted and received signals are coherently processed among the distributed antennas to achieve array gains and spatial interference suppression. 

 Recent works have developed new tools to analyze cell-free massive MIMO performance under practical channel state information \cite{Ngo2017b,Nayebi2017a} and developed distributed signal processing algorithms and resource allocation protocols that are scalable for large-scale deployment \cite{Bjornson2019Aug}. While a central network entity is needed to control a cell-free network, it is desirable to distribute the computations as much as possible over the network infrastructure, to limit the required signaling between the antennas and from the antennas to the central entity. This is a key difference from cloud-RAN where the processing is supposed to be done centrally. 
 
RadioWeaves technology will implement cell-free operation on the distributed topology of intelligent surfaces. It will enable trade-offs between central and distributed processing based on the distributed computing elements that can process antenna signals locally. We see a great potential increase in efficiency of the transmission by letting physical effects of propagation, antenna array response, and transceiver constraints, dictate the resource scheduling in this cell-free networking.

\subsection{Potential of RadioWeaves technology to improve service levels and energy efficiency} 

RadioWeaves have the potential to break through barriers encountered in current networking technologies to meet diverse essential performance metrics for future applications:
\begin{itemize}
    \item \emph{Ultra robust operation.} Real-time and autonomous systems require imperceptible latency at the application level and consistent robust service in the full coverage zone. The latency requirements can be met on the flexible connect-compute fabric that can run applications on resources near the application. Moreover, the dispersed access infrastructure with macro-diversity in the location and orientation of antenna arrays can ensure significantly more robust coverage. We further comment on a potential RadioWeaves-based ultra-robust deployment for a future factory use case.
    \item \emph{Augmented experiences.} There will be a need to scale up capacity to support better and new experiences for large numbers of users in wireless networks. These include high definition video and AR in the context of entertainment, tourism, and gaming. RadioWeaves technology can offer unprecedented throughput to many offered through the cooperation of several large electromagnetic surfaces. The capacity can be scaled up elegantly through the cell-free operation by adding more elements. We illustrate such a deployment for crowd scenarios in Section~\ref{sec:analysis}.
    \item \emph{Energy and bandwidth efficient infrastructure and zero-energy devices.} We identify great opportunities to drastically improve energy efficiency by deploying RadioWeaves technology on three levels: The global energy consumed in ICT networks, the power in the wireless transmission, and the communication energy needs of devices. The energy consumption of ICT networks rapidly rises and raises concerns. Central, cloud-based processing moreover is running into bandwidth bottlenecks. We here propose the more sustainable approach to compute locally whenever and wherever possible in the RadioWeave. This evidently minimizes the total energy consumption of applications, at the expense of local compute engines.   
    
    The wireless power efficiency of RadioWeaves will benefit from the distributed deployment that allows to work in constructive interaction with the physical antenna arrays by exciting them to emit energy in their favorable field of view \cite{Mailloux2005}. The proximity of access points and the extreme focusing capabilities of the large arrays further reduce wireless transmit power. In combination they can bring about improvements of an order of magnitude and higher. 
    
    Devices on their turn will of course also be able to transmit at much lower power thanks to the proximity of access points and great gains of the large arrays. Massive MTC services can be supported through simple random access protocols, as the devices can be resolved by the very large number of antenna elements.  Ultimately communication with zero energy devices can be established through directive backscattering technology implemented on the dispersed antenna arrays.
    \item \emph{Accurate and reliable positioning.} New applications, including autonomously navigating robots and location-aware IoT services, pose requirements on positioning technology far beyond current capabilities, especially in indoor environments. Large intelligent surfaces theoretically provide excellent positioning accuracy. It has been proven however that multi-path components may impact the performance of single array-based systems, particularly near walls \cite{Wilding2018}. A central planar array typically is never well placed to determine all positions withing its coverage reliably. RadioWeaves implementing a distributed topology offer two benefits. First, the probability of LoS will be increased. Secondly, the diversity in both location and orientation of the panels can be taken advantage of to reduce outliers. Hence both the positioning accuracy and reliability are boosted. 
\end{itemize}

\section{Actual deployment of RadioWeaves: anticipating what can (not) be done?}%
\label{sec:analysis}

\subsection{Use cases in need of hyper-connected environments} 

We consider two distinct use cases in need of hyper-connected environments to support future wireless application requirements: a crowd wishing for augmented experiences, and a future factory environment. Both the needs and the propagation conditions are challenging yet different for these two use cases. RadioWeaves technology leveraging on distributed panels with dispersed electronics can be designed to adequately serve the use cases. We one can actively and in a controlled way "generate" favorable propagation conditions to boost service levels. In the following subsections we present high-level characteristics for these two use cases, and anticipate how RadioWeaves technology can make a suitable substrate for the future service needs. We here assume operating frequencies below 10~GHz such that humans and similar sized or smaller objects do not form a blocker for the signals.

\subsection{Boosting capacity: embracing a crowd} 
The capacity of current and emerging wireless networks is insufficient to support a very high number of individual video services and enhanced user experiences to be offered in places where crowds gather. This occurs in large and typically relatively open spaces such as a stadium, on a festival ground, in big halls or auditoriums. 

\begin{figure}[h]
    \centering
    \includegraphics[width=1\linewidth]{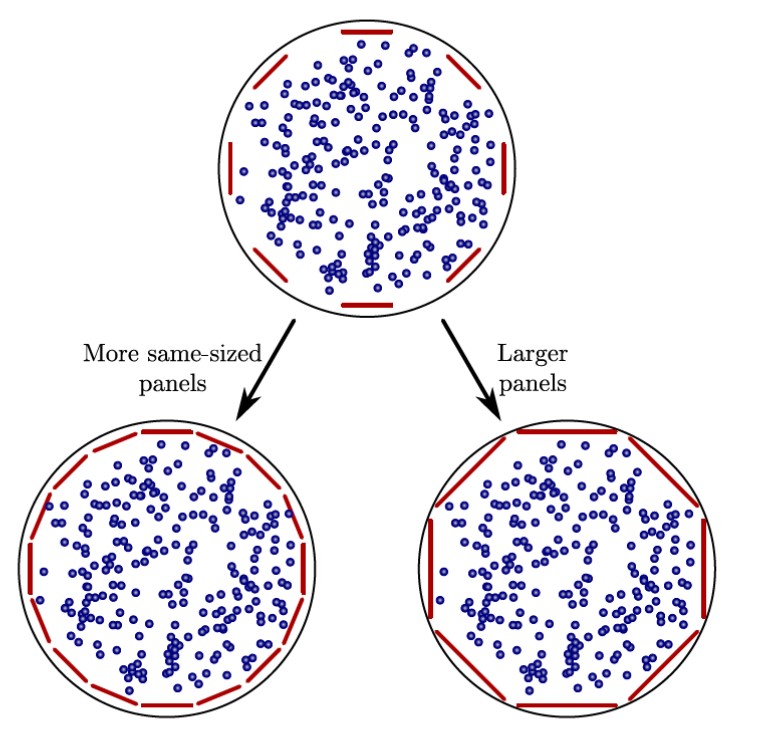}
    \caption{Envisioned RadioWeaves for a crowd scenario, scaling capacity by deploying more or larger panels.}%
    \label{figure:crowd_case}
\end{figure}

A well suited RadioWeaves deployment could "embrace" the crowd from above to create predominantly Line-of-Sight (LoS) propagation conditions, including diffraction, to the many users. Fig.~\ref{figure:crowd_case} shows a potential RadioWeaves panel placement for a crowd scenario. The capacity could be scaled up according to the needs by installing more or larger panels. 

\begin{figure}[h]
    \centering
    \includegraphics[width=1\linewidth]{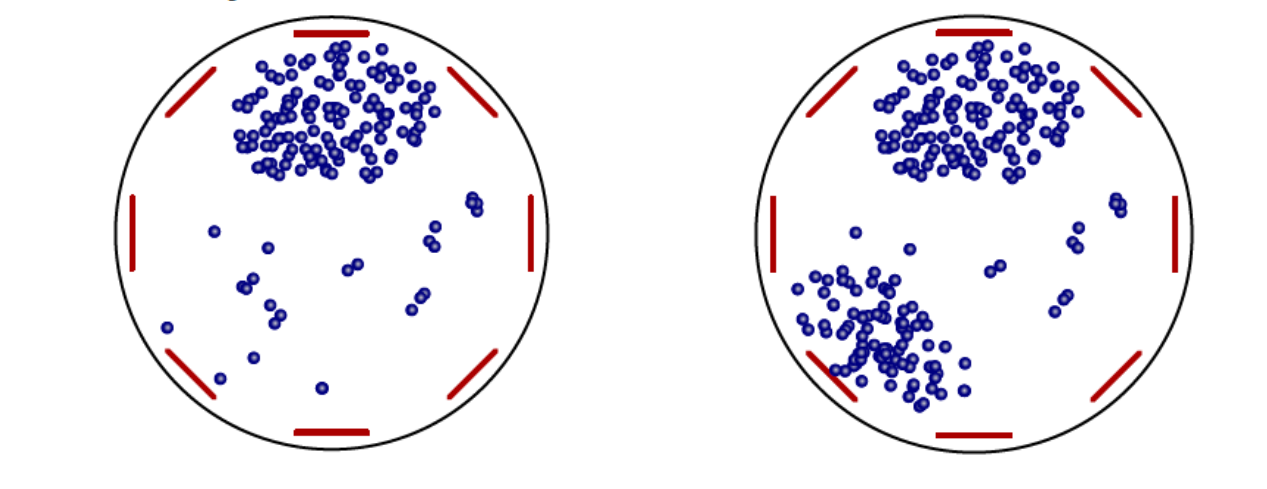}
    \caption{RadioWeave resources can be attributed dynamically to best serve users that may be clustered - consistently or temporarily.}%
    \label{figure:uneven_crowd}
\end{figure}

Fig.~\ref{figure:uneven_crowd} illustrates situations where the users in a crowd are not evenly distributed. The cell-free operation can implement a dynamic allocation of users to (sets of) panels and hence lead to an optimal and adaptive usage of available resources. 

\subsection{Hyper-connected factories: art gallery problem revisited}

In the factory of the future different applications will rely on wireless communication, including ultra reliable and low latency communication and broadband connectivity. Moreover accurate positioning is required for robots, autonomous vehicles and instruments. Ultimately one wants to track all goods as well, based on fully passive devices. A factory environment is typically highly reflective and features many blocking objects. 

\begin{figure}[h]
    \centering
    \includegraphics[width=1\linewidth]{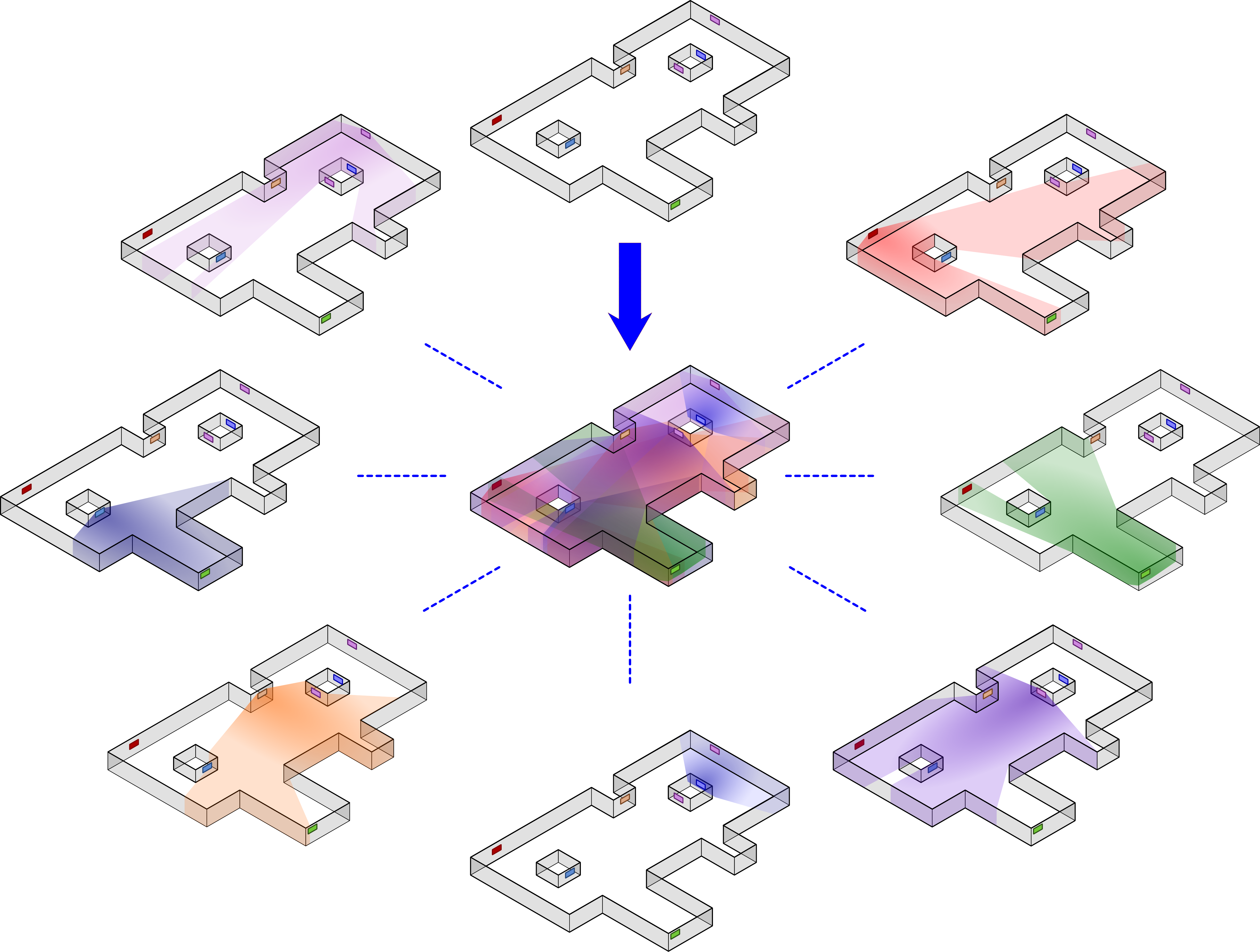}
    \caption{RadioWeaves fit for a hyper-connected factory.}%
    \label{figure:factory_box}
\end{figure}

RadioWeaves can be designed to meet the combined challenging requirements.
The target is "to see every cube cm" by several panels, either via a direct Line-of-Sight (LoS) or via a strong path with a single reflection. This requires solving a 3D art gallery problem \cite{CHVATAL1975}.
Fig.~\ref{figure:factory_box} illustrates a RadioWeaves deployment optimizing reliability in a hyper-connected factory when considered as a 2D problem. The LoS-view of specific panels, shown as components, together cover the entire environment, as illustrated in the central view. 

 The night guide in the art gallery problem for RadioWeaves may get help from "mirrors" presented by strong single reflection paths. Also, objects that manipulate the propagation, for example passive (meta-)surfaces or a strategically positioned  robot, can be entered deliberately into the scene to further increase reliability. Many objects may be in the near field of the arrays when relatively large active surfaces, with respect to both the dimensions of the environment and the wavelength, are installed. This situation opens opportunities for connecting to and positioning of fully passive devices through backscattering systems~\cite{guo2019cooperative}. The realization thereof still poses significant R\&D challenges, as listed in Section~\ref{sec:challenges}.

\section{Impact of deployment constraints: the good, the bad, and the handsome}%
\label{sec:impact}

Actual RadioWeaves deployments will be constrained in a number of different ways. We comment below on how the impact can both be good, bad, and handsome.

\subsection{The good}

The wireless signal propagation is dictated by physics. The fact that the path loss increases quickly with the distance puts a limitation on the signal strength and thus on the sensing and communication performance. Instead of combating the path loss by transmitting with high power from massive MIMO arrays at distant rooftops, RadioWeaves are simply moved close to the users and can therefore achieve the better performance with much lower transmit power.

While a major engineering effort is needed to install and calibrate conventional base stations, the abundance of antennas and computing entities in RadioWeaves makes installation relatively easy---it should become equivalent to hanging up a wallpaper. By spreading out the antennas and dedicated processing units in the vicinity of the antennas, power requirements of these units will be far lower than traditional access points and it will be possible to build them with the low complexity hardware as in terminals.

Antennas are inevitably directive but if they are installed on all the walls in a room (or in the roof), they will naturally provide wireless coverage everywhere in the room. Signals transmitted at low power may not leave the intended coverage zone (particularly if implemented with planar antennas having backplanes), which avoids interference beyond the intended space. Even if a large portion of the RadioWeaves are blocked (e.g., by metallic objects in factory environments), the distributed arrays and huge aperture makes it highly likely that some of the antennas are well-positioned to reach the device. In other words, RadioWeaves provide massive diversity against blockage and other large-scale fading effects. The natural limits will also confine the complexity of scheduling in the cell-free network. 


\subsection{The bad}
 In entirely new buildings or in the frame of drastic refurbishments, the integration of RadioWeaves with good electromagnetic features may be provided in the course of the construction process. On the mid term however, retrofits will be required for installing RadioWeaves infrastructure. Thereby compromises will need to be made to safeguard the functional and aesthetic features on the one hand, and keep installation cost and overhead under control on the other hand. The ad-hoc integration of radiating elements in constructions may seriously affect their radiation characteristics, and introduce undesirable losses.
 
 The interconnection may and will create a significant bottleneck in distributed architectures. The exchange of information will inevitably face bandwidth constraints, as already experienced in massive MIMO systems operating one central array. Moreover, relatively long connections between panels will introduce delays that affect the coherency of the processing. Dedicated calibration and synchronization solutions will need to be developed. In conclusion, integration and distributing access points comes with several significant practical challenges.

\subsection{The handsome}
We envision the RadioWeaves will be ‘invisibly’ embedded in the environment. They will hence scale up wireless services in a non-intrusive way. This is in contrast to current situations where wireless access points often visually constitute disturbing elements, or they are installed out of sight at technically bad locations. This means that RadioWeaves have both technological and aesthetic advantages. Moreover, they offer non-stigmatizing, user-friendly support.

RadioWeave panels, potentially both active and passive, will be installed at locations that are valuable for the networking and practically feasible. The distributed panels---housing both processing and radio capabilities---will be interconnected and powered through a backbone network, which also connects them to a central processing infrastructure.

A test-bed is under construction at the Technology campus in Gent of KU Leuven. It is based on the WikiHouse design approach \cite{Wikihouse}. The walls, ceiling, and floor are modular and composed of ‘tiles’ which will host a gathering of miniature antennas as suggested by the sketch. The architectural concept of creating 3D spaces based on 2D panels aligns nicely with the RadioWeaves architectures.

\begin{figure}[h]
\begin{tabular}{cc}
\begin{subfigure}{.45\linewidth}
  \centering
  \includegraphics[width=\linewidth]{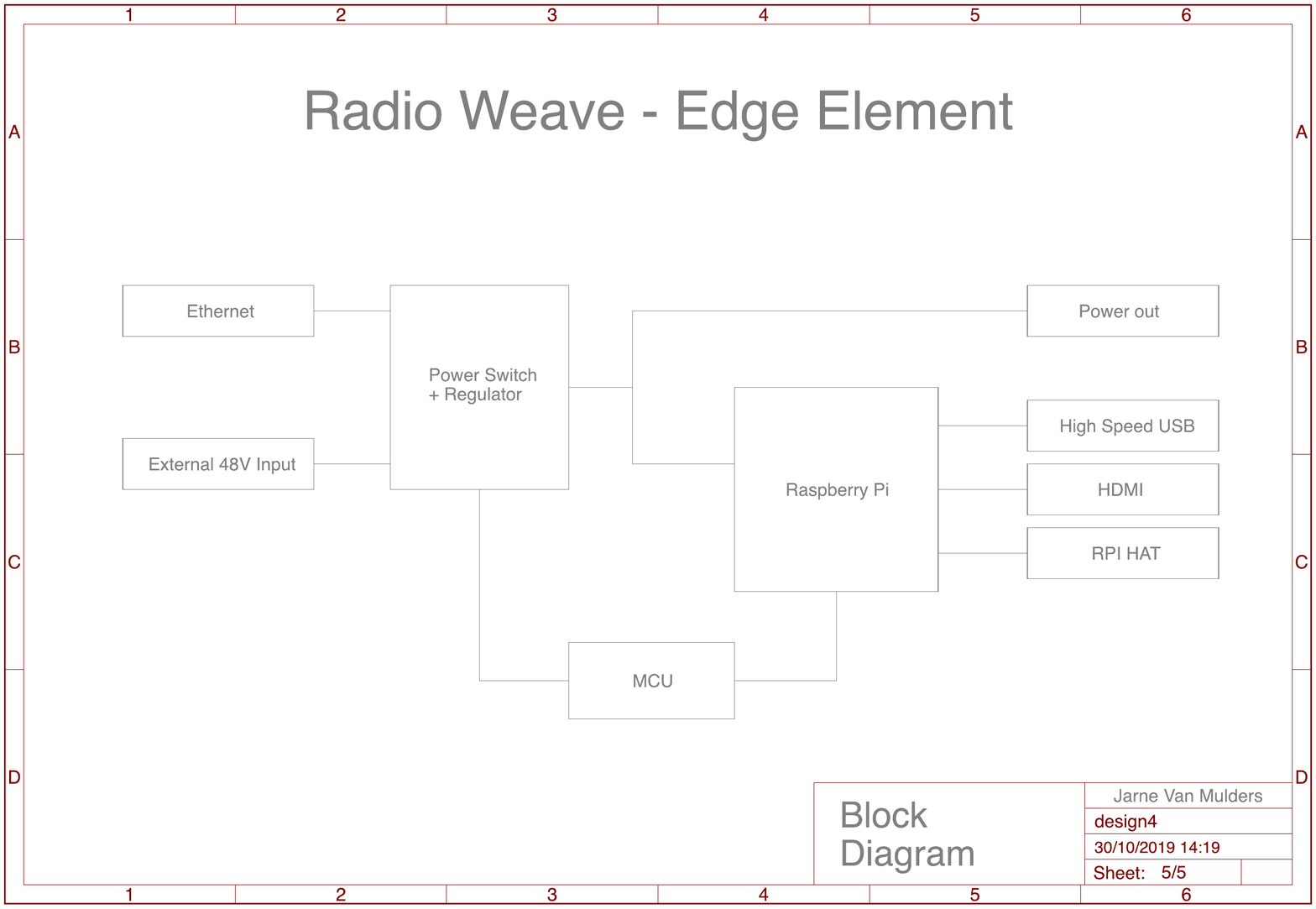}
  \caption{Block scheme}
  \label{fig:schematic}
\end{subfigure}
&
\begin{subfigure}{.45\linewidth}
  \centering
  \includegraphics[width=\linewidth]{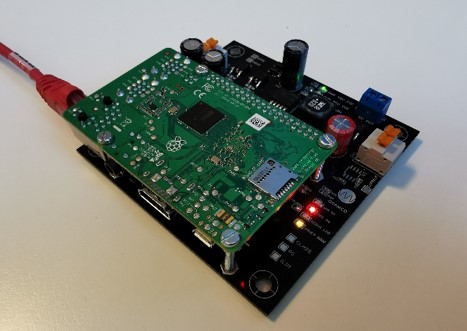}
  \caption{Hardware realisation}
\label{fig:PCB}
\end{subfigure}
\end{tabular}
\caption{Prototype hardware for distributed edge node implementation.}%
    \label{figure:element}
\end{figure}

A prototype board to implement the distributed nodes was developed. Fig.~\ref{figure:element} shows the block diagram of the design and a picture of the hardware.

Both power to and connectivity between the distributed elements are provided via Ethernet cables, which are depicted by the green lines in Fig.~\ref{figure:sketchRadioWeave}. Distributed edge computing devices and access to the cloud are foreseen.

\section{R\&D challenges and opportunities}%
\label{sec:challenges}

We envision a great potential in RadioWeaves technology to create hyper-connected environments offering energy efficient access for heterogeneous services in diverse environments. We have identified several R\&D challenges  to be addressed to achieve this potential, as listed below. It is quite probable that significant others will appear upon progressing insights. Also we see new opportunities, specifically in the backscattering-based connectivity and inclusion of metamaterial-based panels.








\begin{itemize}
\item \emph{Design and placement of distributed radiating elements and  arrays.}
RadioWeaves antennas will be integrated into naturally existing objects. This poses new questions in their design, for example, what materials
are suitable, to what extent can radiating elements be molded into structures or decorative
objects, and so forth. While in conventional antenna design there is great design flexibility, for RadioWeaves there are many pre-determined boundary conditions given by the objects into which the radiating elements are to be integrated, for example, and the presence of structures that will impact radiation patterns.
One specific question is how to arrange both the distributed panels and the radiating elements within the panels appropriately. As mentioned above, many aspects such as mutual coupling \cite{MasourosC2013}, beamwidth, diversity, and practical installation complexity need to be considered.  
Adequate placement of RadioWeaves surfaces to support reliable and energy efficient connectivity and precise and accurate positioning will require solving scheduling problems, as introduced in \ref{sec:analysis}.
\item 
\emph{Electromagnetically favorable processing.}
From a communication theory viewpoint, RadioWeaves build upon the concepts of cell-free massive MIMO and large intelligent surfaces.
Signal processing and resource allocation solutions have so far hardly considered the electromagnetic properties of antenna arrays. The distributed deployment of surfaces opens the opportunity for creating favorable propagation and antenna array interactions.  In addition, adequate signal processing solutions considering distributed versus global processing to optimize performance versus processing and interconnect complexity, are needed.

\item 
\emph{Scalable and adaptive resource allocation.}
A important insight from studies of massive MIMO \cite{Fitzgerald2019} is that power control and allocation is crucial to avoid near-far effects and thereby achieve consistent good quality of service. This insight is key for scheduling users and allocating resources in the new RadioWeaves.  These should be designed to be scalable to large scale deployments. Data-driven and learning approaches can be implemented to smoothly adapt to changes in the environment.


  
 \item
\emph{Partially synchronized operation.}  It is likely that phase coherency may only be maintained in parts of a RadioWeaves installation and not among all radiating elements.
This in turn requires careful analysis and offers opportunities for signal processing algorithm design.

\item 
\emph{Transceiver impairments and hardware friendly transmission schemes.} Hardware non-idealities, most importantly non-linearities and phase noise in the transceivers, can have a serious impact on performance and/or energy efficiency. There is a great interest in identifying in an early phase transmission schemes that relax the transceiver requirements.


\item 
\emph{Integration with backscattering IoT devices.}  
Backscattering is an interesting technology for communication with  low power, ultimately fully passive IoT devices. 
A challenge is that the feasible ranges of communication are quite short.  Using RadioWeaves to interact with backscattering devices has the potential to extend the range, communicate with several zero-energy devices simultaneously, and also support more specific setups, for example where a mobile terminal acts as reader, decoding both a primary signal from the base station and a signal backscattered from an IoT device~\cite{guo2019cooperative}.
 
 \item
\emph{Opportunities offered by passive metamaterials-based panels.} 
Conveniently designed panels using the specific properties of metamaterials may contribute to the creation of favorable propagation conditions and consistent omni-present good service levels by reflecting waves into desired directions \cite{Achouri2015}. They may avoid the need to implement (full) signal processing at and connectivity to all RadioWeave electromagnetic elements.




\end{itemize}

\section{Conclusions and outlook}%
\label{sec:conclusion}


We have presented the concept of RadioWeaves, proposing a new wireless networking technology. This technology will operate a distributed architecture of panels hosting antenna elements as a cell-free infrastructure with unprecedented capabilities. Ultimately RadioWeaves will be realized through the invisibly integration of radiating elements and electronics in the environment. Potential deployment approaches for two challenging use cases, namely crowd scenarios and future factories, have been presented. The impact of actual deployment constraints has been commented on. We have outlined a variety of R\&D challenges to be addressed to pave the way towards adequate realizations of RadioWeaves.

In conclusion we envision RadioWeaves technology has the potential to lift up the energy efficiency and effectiveness of large array-based communication systems. It can serve as one shared substrate to offer consistent ultra robust operation, latencies that are not perceived in the operation of real-time autonomous systems, precise positioning,and support of zero-energy devices. When scaling up the number and/or size of the panels it can also offer superior capacity.

The blended proximity computing-communication paradigm of RadioWeaves is a key asset in the many situations where devices and people in practise physically cooperate ‘in an inner circle’. The approach avoids access to and dependency of global networks and servers when not needed. This contributes to better data privacy and protection thanks to reduction of the "attack surface"clear. Running services in a local 'circle-net-of-things' moreover brings about a global reduction of bandwidth requirements and energy consumption.

\appendices
\section*{Acknowledgment}

The authors thank Sofie Pollin from KU Leuven and Klaus Witrisal from TU Graz for contributing to the RadioWeaves concept, Dimitri Coppens, Geoffrey Ottoy, Jarne Van Mulders, and Gilles Callebaut from KU Leuven who enthusiastically engage in the building of a RadioWeave  instantiation.

{\small
	\bibliography{IEEEabrv,references}{}%
	\bibliographystyle{IEEEtranN}%
}


\end{document}